\documentstyle[preprint,tighten,pra,aps]{revtex}
\begin{document} \draft
%----------------------------------------------------------------------

\title{\Large \bf Sliderule-like property of Wigner's little groups and
cyclic S-matrices for multilayer optics}

\author{Elena Georgieva\footnote{electronic address:
egeorgie@pop500.gsfc.nasa.gov}}
\address{National Aeronautics and Space Administration, Goddard Space
Flight Center, Laser and Electro-Optics Branch, Code 554, Greenbelt,
Maryland 20771}

\author{Y. S. Kim\footnote{electronic address: yskim@physics.umd.edu}}
\address{Department of Physics, University of Maryland, College Park,
Maryland 20742}

\maketitle

\begin{abstract}

It is noted that two-by-two ``S'' matrices in multilayer optics
can be represented by the $Sp(2)$ group whose algebraic property is
the same as the group of Lorentz transformations applicable to
two space-like and one time-like dimensions.  It is noted also
that Wigner's little groups have a sliderule-like property which
allows us to perform multiplications by additions.  It is shown that
these two mathematical properties lead to a cyclic representation
of the S-matrix for multilayer optics, as in the case of $ABCD$
matrices for laser cavities.  It is therefore possible to write
the $N$-layer S-matrix as a multiplication of the $N$ single-layer
S-matrices resulting in the same mathematical expression with one
of the parameters multiplied by $N.$  In addition, it is noted,
as in the case of lens optics, multilayer optics can serve as an
analogue computer for the contraction of Wigner's little groups for
internal space-time symmetries of relativistic particles.

\end{abstract}


\vspace{5mm}

\section{Introduction}\label{intro}
In our previous paper on multilayer optics~\cite{gk01}, it was shown
that the complex two-by-two S-matrix formalism is equivalent to a
two-by-two real matrix representation of the $Sp(2)$ group, which
shares the same algebraic property as the Lorentz group applicable
to two space-like and one time-like dimensions.  This group has
three independent parameters.  It was shown furthermore that, under
certain conditions, one of the off-diagonal elements vanishes,
and the three remaining elements can be computed analytically.
We called this the Iwasawa effect~\cite{gk01}.  In this paper, we
remove those "certain conditions" and achieve the same kind of
simplification for all possible multilayer cases.

Indeed, the group $Sp(2)$ plays the central role in both quantum
and classical optics, including multilayer optics~\cite{sanch99}.
It consists of two-by-two real matrices whose determinant is one.
Each matrix contains at most three independent parameters.  It is
thus a simple matter to multiply two or three matrices.  However,
for multiplication of a large number of matrices presents a new
problem.  The product of those many matrices will also be one
two-by-two matrix with a unit determinant, but how can we calculate
their elements?

For example, let us look at laser cavities.   It consists of a
chain of $N$ identical two-lens systems, where $N$ is the number
of cycles the light beam performs.  The resulting $ABCD$ matrix can
be written as a multiplication of $N$ identical matrices,
but the resulting matrix has the same mathematical form as that
for the single cycle~\cite{bk02}.

Can we then expect a similar cyclic property in multilayer optics?
We have shown in Ref.~\cite{gk01} that the N-dependence can be made
quite transparent if the multilayer S-matrix~\cite{azzam77} is
reduced to the Iwasawa form.  In this paper, we present the cyclic
property for the most general form of multilayers, without the
restriction we imposed in our previous paper~\cite{gk01}.
We shall show that the core of the S-matrix takes the form
\begin{equation}\label{core11}
\pmatrix{\cos\alpha & -\sin\alpha \cr
                   \sin\alpha & \cos\alpha} ,  \quad
\pmatrix{\cosh\beta & \sinh\beta \cr
                    \sinh\beta & \cosh\beta } , \quad
\pmatrix{1 & 0 \cr \gamma & 1} , \quad or \quad
     \pmatrix{1 & \gamma \cr 0 & 1} .
\end{equation}
These matrices form the core of Wigner's little groups applicable
to the internal space-time symmetries of relativistic
particles~\cite{wig39,knp86}. We note here that these matrices
have the following interesting property.

We cannot write $(\cos\alpha_{1} \times \cos\alpha_{2})
= \cos(\alpha_{1} + \alpha_{2}) $ because it is wrong.  However,
in the two-by-two matrix form,
\begin{equation}\label{core55}
\pmatrix{\cos\alpha_{1} & -\sin\alpha_{1} \cr
      \sin\alpha_{1} & \cos\alpha_{1}}
\pmatrix{\cos\alpha_{2} & -\sin\alpha_{2} \cr
      \sin\alpha_{2} & \cos\alpha_{2}}  =
\pmatrix{\cos(\alpha_{1} + \alpha_{2}) & -\sin(\alpha_{1} + \alpha_{2})
\cr \sin(\alpha_{1} + \alpha_{2}) & \cos(\alpha_{1} + \alpha_{2})},
\end{equation}
and similar expressions for the remaining matrices in
Eq.(\ref{core11}).  We call this the sliderule property of Wigner's
little groups.

If they are cycled $N$ times, they take the form
\begin{eqnarray}\label{core33}
&{}&\pmatrix{\cos(N\alpha) & -\sin(N\alpha) \cr
                   \sin(N\alpha) & \cos(N\alpha)} ,  \quad
\pmatrix{\cosh(N\beta) & \sinh(N\beta) \cr
                \sinh(\beta) & (\cosh\beta)} , \nonumber \\[2ex]
&{}& \pmatrix{1 & 0 \cr N\gamma & 1} , \quad and \quad
     \pmatrix{1 & N\gamma \cr 0 & 1} ,
\end{eqnarray}
respectively.  This mathematical instrumentation works for laser
cavity optics~\cite{bk02}.  The question is whether this is
applicable to multilayer optics.

The purpose of this paper is to show that the answer to the above
question is YES.  We note first that the S-matrix consists of
$N$ cycles.  Each cycle consists two phase-shift matrices, one
boundary matrix and its inverse, and this cycle does not take any of
the forms given in Eq.(\ref{core11}) if we start the cycle from the
boundary.  In this paper, we show that it is possible to obtain the
core in the form of Eq.(\ref{core11}) if we start the cycle from
somewhere within one of the media between the two boundaries.

Throughout this paper, we avoid group theoretical languages and rely
on explicit two-by-two matrices with real elements.  However, in so
doing, we are going through an important group theoretical aspect which
became known to us only recently, namely on contractions of Wigner's
little groups.  This aspect was discussed in detail in a recent
paper on lens optics~\cite{bk03}.  Thus, we shall borrow some of the
mathematical identities from that paper.

In addition, in the present paper, we observe that Wigner's little
group has sliderule properties which allow us to convert
multiplications into additions.  This property was noted for one
of the little groups in the paper of Han {\it et al.}  In this
paper, we shall show that all three of the little groups have the
same sliderule property, using Eq.(\ref{core55}).

In Sec.~\ref{formul}, we formulate the problem in terms of the
S-matrix method widely used in multilayer
optics~\cite{azzam77,monzon96,georg97},
and show that the complex S-matrices can be transformed to real
matrices by a conjugate transformation, and thus to the algebra
of the $Sp(2)$ group which is by-now a familiar mathematical
language in optics.
In Sec.~\ref{lorentz}, we import from the literature mathematical
identities useful for the purpose of the present paper.  They
are derivable from Wigner's little groups and their contractions.
In Sec.~\ref{cycle}, using the cyclic property of Eq.(\ref{core33}),
it is shown possible to write the multilayer S-matrix as a
multiplication of the $N$ single-layer S-matrices resulting in the
same mathematical expression with one of the parameters multiplied
by $N.$
In Sec.~\ref{experi}, it is pointed out that the mathematical
identities presented in this paper can be tested experimentally.
We discuss the condition under which the system can achieve the
Iwasawa effect~\cite{gk01}.

\section{Formulation of the Problem}\label{formul}
It was noted our previous paper that one cycle in $N$-layer optics
starts with the boundary matrix of the form~\cite{monzon00}
\begin{equation}\label{bd11}
B(\eta) = \pmatrix{\cosh(\eta/2) & \sinh(\eta/2) \cr \sinh(\eta/2) &
\cosh(\eta/2) } ,
\end{equation}
which describes the transition from $medium~2$ to $medium~1$, taking
into account both the transmission and reflection of the beam.  As
the beam goes through the $medium~1$, the beam undergoes the phase
shift represented by the matrix
\begin{equation}\label{ps11}
P(\phi_1) = \pmatrix{e^{-i\phi_1/2} & 0 \cr 0 & e^{i\phi_1/2}} .
\end{equation}
When the wave hits the surface of the second medium, the corresponding
matrix is
\begin{equation}\label{bd22}
B(-\eta) = \pmatrix{\cosh(\eta/2) & -\sinh(\eta/2) \cr -\sinh(\eta/2) &
\cosh(\eta/2) } ,
\end{equation}
which is the inverse of the matrix given in Eq.(\ref{bd11}).
Within the second medium, we write the phase-shift matrix as
\begin{equation}\label{ps22}
P(\phi_2) = \pmatrix{e^{-i\phi_2/2} & 0 \cr 0 & e^{i\phi_2/2}} .
\end{equation}
Then, when the wave hits the first medium from the second, we have
to go back to Eq.(\ref{bd11}).
Thus, one cycle consists of
\begin{eqnarray}
&{}& M_1 = \pmatrix{\cosh(\eta/2) &
\sinh(\eta/2) \cr \sinh(\eta/2) & \cosh(\eta/2) }
\pmatrix{e^{-i\phi_1/2} & 0 \cr 0 & e^{i\phi_1/2}}
\pmatrix{\cosh(\eta/2) & -\sinh(\eta/2) \cr -\sinh(\eta/2) & \cosh(\eta/2) }
\nonumber \\[2ex]
&{}& \hspace{20mm}
\times \pmatrix{e^{-i\phi_2/2} & 0 \cr 0 & e^{i\phi_2/2}} .
\end{eqnarray}
This matrix contains complex numbers, but we are interested in carrying
out calculations with real matrices.  This can be done if we make the
following conjugate transformation~\cite{gk01}

Let us next  consider the matrix
\begin{equation}
C = {1 \over 2} \pmatrix{1 & 1 \cr -1 & 1} \pmatrix{1 & i \cr i & 1}
 = {1 \over \sqrt{2}} \pmatrix{e^{i\pi/4} &  e^{i\pi/4} \cr
-e^{-i\pi/4} & e^{-i\pi/4}} .
\end{equation}
Then we have shown in our previous paper that
\begin{equation}\label{conju11}
M_{2} = C~M_{1}~C^{-1} ,
\end{equation}
with
\begin{eqnarray}\label{core22}
&{}& M_{2} = \pmatrix{e^{\eta/2} & 0 \cr 0 & e^{-\eta/2} }
\pmatrix{\cos(\phi_1/2) & -\sin(\phi_1/2) \cr
\sin(\phi_1/2) &  \cos(\phi_1/2)}
\pmatrix{ e^{-\eta/2} & 0 \cr 0 & e^{\eta/2}}
\nonumber \\[2ex]
&{}& \hspace{26mm} \times
\pmatrix{\cos(\phi_2/2) & -\sin(\phi_2/2) \cr
\sin(\phi_2/2) &  \cos(\phi_2/2) }  .
\end{eqnarray}

The conjugate transformation of Eq.(\ref{conju11}) changes
the boundary matrix $B(\eta)$ of Eq.(\ref{bd11}) to a squeeze matrix
\begin{equation}\label{sq11}
S(\eta) = \pmatrix{ e^{\eta/2} & 0 \cr 0 & e^{-\eta/2} } ,
\end{equation}
and the phase-shift matrices $P(\phi_1)$ of Eq.(\ref{ps11}) and
Eq.(\ref{ps22}) to rotation matrices
\begin{equation}\label{rot22}
R(\phi_{i}) = \pmatrix{\cos(\phi_{i}/2) & -\sin(\phi_{i}/2) \cr
\sin(\phi_{i}/2) &  \cos(\phi_{i}/2) } ,
\end{equation}
with $i = 1,~2$.

Indeed, the matrices $M_1$ and $M_2$ can be written as
\begin{eqnarray}
&{}& M_1 = B(\eta) P(\phi_1) B(-\eta) P(\phi_2) , \nonumber \\[2ex]
&{}& M_2 = S(\eta) R(\phi_1) S(-\eta) R(\phi_2) .
\end{eqnarray}
The matrix $M_2$ can be obtained from $M_1$ by the conjugate
transformation in Eq.(\ref{conju11}).  Conversely, $M_1$ can be
obtained from $M_2$ through the inverse conjugate transformation:
\begin{equation}\label{conju55}
M_{1}  = C^{-1}~M_{2}~C .
\end{equation}

In addition, the conjugate transformations have the following properties.
\begin{equation}\label{conju33}
\left(M_2 \right)^{N} = C~\left(M_1\right)^{N}~C^{-1} , \quad
\left(M_1 \right)^{N} = C^{-1}~\left(M_2\right)^{N}~C .
\end{equation}
Thus, we can study $M_2$ in order to study $M_1$.  The advantage of
$M_2$ is that it consists of real matrices.  The group of these matrices
is called $Sp(2)$ which is like (isomorphic) the Lorentz group
applicable to three space and one time dimensions.  This group contains
very rich group theoretical contents including those of Wigner's little
groups.  We intend to study $M_2$ in terms of those little groups.

The problem is that $M_2$ takes a simple form, and
$\left(M_{2}\right)^{2}$ is manageable, but we cannot predict what form
$\left(M_{2}\right)^{N}$ takes.  In this paper, we shall construct the
core matrix of the form of Eq.(\ref{core11}) for multilayer optics.
Then, as we can see in Eq.(\ref{core33}), the chain effect is
straight-forward.  We shall calculate $\left(M_{2}\right)^{N}$ first
and then $\left(M_{1}\right)^{N}$.

\section{Mathematical Identities from the Lorentz Group}\label{lorentz}
Wigner's little groups were formulated for internal space-time
symmetries of relativistic particles~\cite{wig39,knp86}.  However,
they produced many mathematical identities useful in other branches
of physics, including classical layer optics which depends heavily
on two-by-two matrices.  The correspondence between the two-by-two
and four-by-four representations of the Lorentz group has been
repeatedly discussed in the literature~\cite{gk01,bk02,bk03}.  In the
two-by-two representation, we write the rotation matrix around the
$y$ axis as
\begin{equation}\label{b11}
\pmatrix{\cos(\phi/2) & -\sin(\phi/2) \cr \sin(\phi/2) & \cos(\phi/2) } ,
\end{equation}
and the boost matrices along the $z$ and $x$ axes as
\begin{equation}\label{b22}
\pmatrix{e^{\eta/2}  & 0 \cr 0 & e^{-\eta/2}}, \qquad
\pmatrix{\cosh\lambda & \sinh\lambda \cr \sinh\lambda & \cosh\lambda} ,
\end{equation}
respectively.  We shall use only these three matrices in this paper.

We use the following identity which Baskal and Kim
introduced recently in their paper on lens optics and group
contractions~\cite{bk03,hk88}.
\begin{eqnarray}\label{eqn11}
&{}&\pmatrix{e^{\eta/2}  & 0 \cr 0 & e^{-\eta/2}}
\pmatrix{\cos(\phi/2) & -\sin(\phi/2) \cr
 \sin(\phi/2) & \cos(\phi/2) }
\pmatrix{e^{-\eta/2}  & 0 \cr 0 & e^{\eta/2}}   \nonumber \\[3ex]
&{}& = \pmatrix{\cos(\theta/2) & -\sin(\theta/2) \cr \sin(\theta/2) & \cos(\theta/2)}
\pmatrix{\cosh\lambda & \sinh\lambda \cr \sinh\lambda & \cosh\lambda}
\pmatrix{\cos(\theta/2) & -\sin(\theta/2) \cr \sin(\theta/2) & \cos(\theta/2)} ,
\end{eqnarray}
with
\begin{eqnarray}\label{eqn22}
&{}& \cos(\phi/2) = \cosh\lambda \cos\theta, \nonumber \\[3ex]
&{}& e^{2\eta} = {\cosh\lambda \sin\theta + \sinh\lambda  \over
     \cosh\lambda \sin\theta - \sinh\lambda} .
\end{eqnarray}
The left-hand side of the above expression is one rotation matrix
sandwiched by one boost matrix and its inverse, while the right-hand
side consists of one boost matrix sandwiched between two identical
rotation matrices.

The left-hand side of Eq.(\ref{eqn11}) is the same as the first
three matrices of the core matrix $M_{2}$ given in Eq.(\ref{core22}).
However, the fourth matrix is a rotation matrix.  Since one-rotation
matrix multiplied by another rotation matrix is still a rotation matrix,
the core matrix $M_{2}$ is one boost matrix sandwiched between two
different rotation matrices.  Thus, the problem is to find a
transformation which will make those two rotation matrices the same,
and go back to the form of the left-hand side of Eq.(\ref{eqn11}).
We shall come  back to this problem in Sec.~\ref{cycle}.

If we complete the matrix multiplications of both side, the result is
\begin{equation}\label{eqn33}
\pmatrix{\cos(\phi/2) & -e^{\eta} \sin(\phi/2) \cr
 e^{-\eta} \sin(\phi/2) & \cos(\phi/2)}
= \pmatrix{\cosh\lambda \cos\theta
    & -(\cosh\lambda \sin\theta + \sinh\lambda) \cr
 \cosh\lambda \sin\theta - \sinh\lambda  &
 \cosh\lambda \cos\theta } .
\end{equation}
Then we can write $\phi$ and $\eta$ in terms of
$\lambda$ and $\theta$ as given in Eq.(\ref{eqn22}).
The parameters $\lambda$ and $\theta$ can be written in terms of
$\phi$ and $\eta$ as
\begin{eqnarray}
&{}& \cosh\lambda =
 (\cosh\eta)\sqrt{1 - \cos^{2}(\phi/2)\tanh^{2}\eta}, \nonumber
\\[2ex]
&{}& \cos\theta = { \cos(\phi/2) \over (\cosh\eta)
      \sqrt{1 - \cos^{2}(\phi/2)\tanh^{2}\eta} } .
\end{eqnarray}

The above relation is valid only for
$(\cosh\lambda \sin\theta/2 - \sinh\lambda)$ is positive.  If it is
negative, the left-hand side of the above expression should be
\begin{eqnarray}\label{eqn44}
&{}& \pmatrix{e^{\eta/2}  & 0 \cr 0 & e^{-\eta/2}}
\pmatrix{\cosh(\chi/2) & -\sinh(\chi/2) \cr
 - \sinh(\chi/2) & \cosh(\chi/2)}
\pmatrix{e^{-\eta/2}  & 0 \cr 0 & e^{\eta/2}}  \nonumber \\[2ex]
&{}& = \pmatrix{\cosh(\chi/2) & -e^{\eta} \sinh(\chi/2) \cr
- e^{-\eta} \sinh(\chi/2) & \cosh(\chi/2) } ,
\end{eqnarray}
with
\begin{eqnarray}
\cosh(\chi/2) = \cosh\lambda \cos\theta, \nonumber \\[2ex]
e^{2\eta} = {\cosh\lambda \sin\theta + \sinh\lambda  \over
 \sinh\lambda - \cosh\lambda \sin\theta } .
\end{eqnarray}
Conversely, $\lambda$ and $\theta$ can be written in terms of
$\chi$ and $\eta$ as
\begin{eqnarray}
&{}& \cosh\lambda =
(\cosh\eta)\sqrt{\cosh^{2}(\chi/2) - \tanh^{2}\eta} \, , \nonumber
\\[2ex]
&{}& \cos\theta = {\cosh(\chi/2) \over
 (\cosh\eta)\sqrt{\cosh^{2}(\chi/2) - \tanh^{2}\eta} }  .
\end{eqnarray}

An interesting case is when $\sinh\lambda - \cosh\lambda \sin\theta$
becomes zero, and $\eta$ becomes very large.  If we insist that
\begin{equation}
e^{\eta}\sin(\phi/2) = u,
\end{equation}
remain finite, then $\phi/2$ must become very small.  On the right-hand
side,
\begin{equation}
u = 2\sinh\lambda, \quad with \quad \sin\theta = \tanh\lambda .
\end{equation}
The net result is that both sides take the form
\begin{equation}\label{eqn55}
\pmatrix{1 & -2\sinh\lambda \cr 0 & 1}.
\end{equation}

In their recent paper~\cite{bk03}, Kim and Baskal studied in detail
the transition
from Eq.(\ref{eqn33}) to Eq.(\ref{eqn44}) through Eq.(\ref{eqn55}),
and showed that the one-lens camera goes through this transition as
we try to focus the image.  Mathematically, the system goes through
group contraction processes.  In the present paper, we show that
the same contraction process can be achieved in multilayer optics.

\section{Cyclic Representation of the S Matrix}\label{cycle}
It was noted in Sec.~\ref{formul} that each cycle consists of
\begin{equation}\label{cycle33}
  \left(S R_{1} S^{-1} R_{2}\right) ,
\end{equation}
with
\begin{equation}
R_1 = R(\phi_1), \qquad R_2 = R(\phi_2) ,
\end{equation}
of Eq.(\ref{rot22}) respectively.  The squeeze matrix $S$ is
given in Eq.(\ref{sq11}).
For the layer consisting of $N$ cycles, let us consider the chain
\begin{equation}\label{chain11}
M_{2}^{N} = \left(S R_{1} S^{-1} R_{2}\right)
  \left(S R_{1} S^{-1} R_{2}\right)
  \left(S R_{1} S^{-1} R_{2}\right)
   .... \left(S R_{1} S^{-1} R_{2}\right) .
\end{equation}

According to Eq.(\ref{eqn11}), we can now write $S R_{1} S^{-1}$
in the above expression as
\begin{equation}
S R_{1} S^{-1} = R_{3}~X~R_{3} ,
\end{equation}
with
\begin{equation}
 R_{3} = \pmatrix{\cos(\phi_{3}/2) & -\sin(\phi_{3}/2) \cr
            \sin(\phi_{3}/2) &  \cos(\phi_{3}/2) } ,  \qquad
 X = \pmatrix{\cosh\lambda & \sinh\lambda \cr
                    \sinh\lambda &  \cosh\lambda } ,
\end{equation}
and
\begin{eqnarray}
&{}& \cosh\lambda = (\cosh\eta)\sqrt{1 -
 \cos^{2}(\phi_1/2)\tanh^{2}\eta}, \nonumber \\[2ex]
&{}& \cos\phi_{3} = { \cos(\phi_1/2) \over (\cosh\eta)
      \sqrt{1 - \cos^{2}(\phi_1/2)\tanh^{2}\eta} } .
\end{eqnarray}
The parameters $\lambda$ and $\phi_{3}$ are determined from
$\eta$ and $\phi_{1}$ which are the input parameters from the
optical properties of the media.

The chain of Eq.(\ref{chain11}) becomes
\begin{equation}\label{chain22}
M_{2}^{N} = \left(R_{3} X R_{3} R_{2} \right)
            \left(R_{3} X R_{3} R_{2}\right)
            \left(R_{3} X R_{3} R_{2}\right) ....
            \left(R_{3} X R_{3} R_{2}\right) .
\end{equation}
Let us next introduce the rotation matrix $R(\alpha)$ as
\begin{equation}
R(\alpha) = \left(R_{2}\right)^{1/2} R_{3} ,
\end{equation}
with
\begin{equation}\label{alpha}
\alpha = \phi_{3} + {1 \over 2} \phi_{2} ,
\end{equation}
where $\phi_{2}$ is an input parameter.  Since $\phi_{3}$ is determined
by $\eta$ and $\phi_{1}$, the rotation angle $\alpha$
is determined by the three input parameters, namely $\eta$, $\phi_{1}$,
and $\phi_{2}$.

In terms of $R = R(\alpha)$, the chain of Eq.(\ref{chain22})
becomes
\begin{equation}\label{chain33}
M_{2}^{N} = R_{3} R^{-1}(R X R) (R X R) (R X R) ....
            (R X R) R^{-1} R_{3} R_{2} .
\end{equation}\label{chain44}
Since $R_{3} R^{-1} = R_{2}^{-1/2}$ and
$R^{-1} R_{3} R_{2} = R_{2}^{1/2}$ from Eq.(\ref{alpha}),
\begin{equation}\label{chain55}
M_{2}^{N} = \left(R_{2}\right)^{-1/2}(R X R)
(R X R)(R X R)
..... (R X R)(R_{2})^{1/2} .
\end{equation}
According to Eq.(\ref{eqn11}) and Eq.(\ref{eqn33}), we can now
write $R X R $ as
\begin{equation}\label{core77}
R X R = \pmatrix{\cosh\lambda \cos\alpha
    & -(\cosh\lambda \sin\alpha + \sinh\lambda) \cr
 \cosh\lambda \sin\alpha - \sinh\lambda  &
 \cosh\lambda \cos\alpha } .
\end{equation}
According to the formulas given in Sec.~\ref{lorentz},
especially Eq.(\ref{eqn11}),
$R X R$ can also be written as
\begin{equation}
R X R = Z A Z^{-1} ,
\end{equation}
with
\begin{equation}
Z = \pmatrix{e^{\xi/2} & 0 \cr 0 & e^{-\xi/2}} .
\end{equation}
Now the two-by-two matrix $A$ can take one of the following forms.

If the off-diagonal elements of the matrix  of Eq.(\ref{core77})
has opposite signs, the $A$ matrix becomes
\begin{equation}\label{a11}
A = \pmatrix{\cos(\phi/2) & -\sin(\phi/2) \cr
    \sin(\phi/2) & \cos(\phi/2) } ,
\end{equation}
with
\begin{eqnarray}
&{}& \cos(\phi/2) = \cosh\lambda \cos\alpha, \nonumber \\[3ex]
&{}& e^{2\xi} = {\cosh\lambda \sin\alpha + \sinh\alpha  \over
     \cosh\lambda \sin\alpha - \sinh\lambda} .
\end{eqnarray}
If, on the other hand, the off-diagonal elements of the matrix
$R X R$ have the same sign, the matrix $A$ should be written as
\begin{equation}\label{a22}
A = \pmatrix{\cosh(\chi/2) & -\sinh(\chi/2) \cr
-\sinh(\chi/2) & \cosh(\chi/2) } ,
\end{equation}
with
\begin{eqnarray}
\cosh(\chi/2) = \cosh\lambda \cos\alpha, \nonumber \\[2ex]
e^{2\xi} = {\cosh\lambda \sin\alpha + \sinh\lambda  \over
 \sinh\lambda - \cosh\lambda \sin\alpha } .
\end{eqnarray}

We note from Eq.(\ref{a11}) and Eq.(\ref{a22}) that the matrix
$A$ takes circular or hyperbolic forms depending on the sign of
the lower-left element of Eq.(\ref{core77}) which is
\begin{equation}\label{lleft}
\sinh\lambda - (\sin\alpha)\cosh\lambda  ,
\end{equation}
and note that this expression can become from a positive to negative
number continuously as the parameters $\lambda$ and $\alpha$ vary.
These two parameters are determined from the reflection and
transmission properties of the media.

While expression of Eq.(\ref{lleft}) makes the continuous transition,
it has to go through zero.  If it vanishes,
\begin{equation}\label{a33}
RXR = \pmatrix{1 & -2\sinh\lambda \cr 0 & 1} .
\end{equation}
The transition of $A$ from Eq.(\ref{a11}) to Eq.(\ref{a22}) through
this process has been discussed in detail in Ref.~\cite{bk03} in
connection with the contraction of Wigner's little groups.

As we noted in Sec.~\ref{formul}, the matrix $A$ has the desired
cyclic property.  Thus,
\begin{equation}
M_{2}^{N} = \left(R_{2}\right)^{-1/2} \left[\left(Z A Z^{-1}\right)
\left(Z A Z^{-1}\right) \left(Z A Z^{-1}\right)
 ... \left(Z A Z^{-1}\right) \right]
\left(R_{2}\right)^{1/2} .
\end{equation}
Consequently
\begin{equation}
M_{2}^{N} = \left(R_{2}\right)^{-1/2}\left[ Z~A^{N}~Z^{-1} \right]
 \left(R_{2}\right)^{1/2} .
\end{equation}

If $A$ takes the form of Eq.(\ref{a11}),
\begin{equation}\label{a11n}
A^{N} = \pmatrix{\cos(N\phi/2) & -\sin(N\phi/2)
\cr \sin(N\phi/2) & \cos(N\phi/2)} .
\end{equation}
For $A$ given in Eq.(\ref{a22}),
\begin{equation}\label{a22n}
A^{N} = \pmatrix{\cosh(N\chi/2) & -\sinh(N\chi/2) \cr
-\sinh(N\chi/2) & \cosh(N\chi/2) } .
\end{equation}
As Eq.(\ref{a33}),
\begin{equation}\label{a33n}
(RXR)^{N} = \pmatrix{1 & -2N\sinh\lambda \cr 0 & 1} .
\end{equation}

Then the calculation of $\left(M_{2}\right)^{N}$ for the $N$-layer
case is straight-forward.  We can now compute the matrix
$\left(M_{1}\right)^{N}$ using the conjugate transformation of
Eq.(\ref{conju33}).  Let us write our result in two-by-two matrices:
\begin{eqnarray}
&{}& M_{2}^{N} = \left[\pmatrix{\cos(\phi_{2}/4) &
 - \sin(\phi_{2}/4) \cr \sin(\phi_{2}/4) & \cos(\phi_{2}/4)}
  \pmatrix{e^{\xi/2} & 0 \cr 0 & e^{-\xi/2}} \right]
  \pmatrix{\cos(N\phi/2) & -\sin(N\phi/2)
  \cr \sin(N\phi/2) & \cos(N\phi/2)}
  \nonumber \\[2ex]
&{}& \hspace{10mm} \times
     \left[ \pmatrix{e^{-\xi/2} & 0 \cr 0 & e^{\xi/2}}
      \pmatrix{\cos(\phi_{2}/4) & \sin(\phi_{2}/4) \cr
       -\sin(\phi_{2}/4) & \cos(\phi_{2}/4)} \right],
\end{eqnarray}
for $A$ of Eq.(\ref{a11}).  For $A$ of Eq.(\ref{a22}),
\begin{eqnarray}
&{}& M_{2}^{N} = \left[\pmatrix{\cos(\phi_{2}/4) & -\sin(\phi_{2}/4) \cr
      \sin(\phi_{2}/4) & \cos(\phi_{2}/4)}
  \pmatrix{e^{\xi/2} & 0 \cr 0 & e^{-\xi/2}} \right]
  \pmatrix{\cosh(N\chi/2) & -\sinh(N\chi/2)  \cr
  -\sin(N\chi/2) & \cos(N\chi/2)}  \nonumber \\[2ex]
&{}& \hspace{10mm} \times
  \left[   \pmatrix{e^{-\xi/2} & 0 \cr 0 & e^{\xi/2}}
      \pmatrix{\cos(\phi_{2}/4) & \sin(\phi_{2}/4) \cr
       -\sin(\phi_{2}/4) & \cos(\phi_{2}/4)} \right].
\end{eqnarray}

If the lower-left element given in Eq.(\ref{lleft}) vanishes, we
have to go back to Eq.(\ref{chain55}) and Eq.(\ref{a33}), and
write
\begin{equation}
 M_{2}^{N} = \pmatrix{\cos(\phi_{2}/4) & -\sin(\phi_{2}/4) \cr
      \sin(\phi_{2}/4) & \cos(\phi_{2}/4)}
    \pmatrix{1 & -2N\sinh\lambda   \cr 0 & 1}
      \pmatrix{\cos(\phi_{2}/4) & \sin(\phi_{2}/4) \cr
       -\sin(\phi_{2}/4) & \cos(\phi_{2}/4)} .
\end{equation}

As we noted in Sec.~\ref{formul}, we use $M_{2}$ and $M_{2}^{N}$ for
mathematical convenience.  In the real world, we have to use
$M_{1}$ and $M_{1}^{N}$.  It is not difficult to write this expression
using the conjugate transformation of Eq.(\ref{conju55}).  It can be
written as
\begin{eqnarray}
&{}& M_{1}^{N} = \left[\pmatrix{e^{-i\phi_{2}/4} & 0 \cr 0 &
  e^{i\phi_{2}/4}}
  \pmatrix{\cosh(\xi/2) & \sinh(\xi/2)  \cr
  \sinh(\xi/2) & \cosh(\xi/2)} \right]
  \pmatrix{e^{-iN\phi/2} & 0 \cr 0 & e^{iN\phi/2}}
  \nonumber \\[2ex]
&{}& \hspace{7mm}
  \times
 \left[\pmatrix{\cosh(\xi/2) & -\sinh(\xi/2) \cr -\sinh(\xi/2)
  & \cosh(\xi/2)}
 \pmatrix{e^{i\phi_{2}/4} & 0 \cr 0 & e^{-i\phi_{2}/4}}\right] .
\end{eqnarray}
if $A$ takes the form of Eq.(\ref{a11}) with a positive value
of Eq.(\ref{lleft}).  If it takes the form of
Eq.(\ref{a22}) with a negative value of Eq.(\ref{lleft}),
\begin{eqnarray}
&{}& M_{1}^{N} = \left[\pmatrix{e^{-i\phi_{2}/4} & 0 \cr 0 &
  e^{i\phi_{2}/4}}
  \pmatrix{\cosh(\xi/2) & \sinh(\xi/2)  \cr
      \sinh(\xi/2) & \cosh(\xi/2)} \right]
 \pmatrix{\cosh(N\chi/2) & i \sinh(N\chi/2) \cr
 -i \sinh(N\chi/2) & \cosh(N\chi/2)}  \nonumber \\[2ex]
&{}& \hspace{7mm} \times
 \left[\pmatrix{\cosh(\xi/2) & -\sinh(\xi/2) \cr -\sinh(\xi/2)
  & \cosh(\xi/2)}
 \pmatrix{e^{i\phi_{2}/4} & 0 \cr 0 & e^{-i\phi_{2}/4}}\right]  .
\end{eqnarray}
If the expression of Eq.(\ref{lleft}) vanishes,
\begin{equation}
 M_{1}^{N} = \pmatrix{e^{-i\phi_{2}/4} & 0 \cr 0 &  e^{i\phi_{2}/4}}
  \pmatrix{ 1 - iN\sinh\lambda  & iN\sinh\lambda  \cr
  -iN\sinh\lambda  & 1 + iN\sinh\lambda }
 \pmatrix{e^{i\phi_{2}/4} & 0 \cr 0 & e^{-i\phi_{2}/4}} .
\end{equation}

This is not yet the S-matrix.  The first and the last layers have
boundaries with air or the third medium.  It is straight-forward to
take these boundary conditions into consideration.  This procedure was
discussed in detail in our previous paper~\cite{gk01}.

\section{Experimental Possibilities}\label{experi}
The variables for the S-matrix given in Secs.~\ref{lorentz} and
\ref{cycle} are determined by the optical parameters, namely the
two phase-shifts and one reflection/transmission coefficient.
The combinations of these three variables will determine the
form of the S-matrix, which may take three different forms.

We note first the that $N$-dependence of the S-matrix comes from
the form of $A$ matrix or the $RXR$ matrix of Eq.(\ref{core77}).
If the optical parameters are in such a way that the $A$ matrix takes
the form of Eq.(\ref{a11}), the elements of the $A^{N}$ matrix
of Eq.(\ref{a11n}) are bounded and oscillating functions of $N$.
If $A$ takes the form of Eq.(\ref{a22}), the $A^{N}$ matrix becomes
Eq.(\ref{a22n}).  The elements of this matrix are not bounded as
$N$ becomes large.  Thus, in the real world, $N$-layers can have
two different types depending on the form of $A$.

In addition, the optical layers can satisfy the condition that
the expression of Eq.(\ref{lleft}) be zero:
\begin{equation}
\sinh\lambda - (\sin\alpha)\cosh\lambda  = 0 .
\end{equation}
Then the $RXR$ matrix takes the form of Eq.(\ref{a33}), and
the $N$ dependence is linear.  This case can be tested as the
optical parameters are varied from positive values of Eq.(\ref{lleft})
to a positive value through zero.  This condition does not
depend on $N$.  We have discussed a similar case in our previous
paper~\cite{gk01}.

In their recent paper~\cite{bk03}, Baskal and Kim noted the same
transition process for one-lens optics.  They noted that
the camera focusing mechanism corresponds to contraction of
Wigner little groups.  It is interesting to note that the same
contraction mechanism exists in $N$-layer optics.

\section*{Concluding Remarks}

Based on Wigner's little groups, we have developed an algebraic method
which allows us to study the cyclic properties of two-by-two S-matrices
for multilayer optics.  Starting from the single-layer S-matrix, it is
possible to write the $N$-layer matrix by multiplying one of the
parameters by $N$.   The $N$-dependence is therefore transparent.

This is possible because the core matrices of the Wigner's little
groups have a sliderule property which allows us to perform
multiplications by additions, as noted in Eq.(\ref{core55}).
This property is an important element in computer designs.

As was noted in Ref.~\cite{bk03}, the transition from Eq.(\ref{a11})
to Eq.(\ref{eqn22}) corresponds to camera focusing in one-lens optics.
From the mathematical point of view, it corresponds to the contraction
and expansion of the little groups.  From the geometrical point of
view, this corresponds to transformation from a circle to hyperbola.
It is interesting to note that we can perform these operations also
in multilayer optics.  Indeed, as in the case of
lens optics~\cite{bk03}, multilayer optics can serve as an analogue
computer for group contractions.

The correspondence between the Lorentz group $O(3,1)$ and $SL(2,c)$,
the group of two-by-two unimodular matrices, is well known.  Since
most of the matrices in ray optics are two-by-two, the Lorentz group
is becoming the major language in this field.  Ray optics is the
backbone of future technology, and optical devices such as
polarizers, lenses, interferometers, mutilayers, all
speak the language of the Lorentz group.  Thus, it is possible for
the Lorentz group to play computational roles in future
generations of computers.

It is a prevailing view in physics, especially in optics, that
group theory is only for studying symmetries and not useful for
computational purposes.  Indeed, we do not need group theory
to carry out matrix multiplications given in this paper, and we
started only with three matrices given in Eq.(\ref{b11}) and
Eq.(\ref{b22}).  However, are going through some important
theorems in group theory while going through the simple matrix
algebras given in this paper.  We choose not to elaborate on
this point.


\begin{thebibliography}{99}

\bibitem{gk01}
 E. Georgieva and Y. S. Kim, Phys. Rev. E {\bf 64}, 026602 (2001).

\bibitem{sanch99}
 J. J. Monzon and L. L. Sanchez-Soto, Opt. Commun. {\bf 162}, 1 (1999);
 J. J. Monzon and L. L. Sanchez-Soto, Phys. Lett. A {\bf 262}, 18 (1999);
 J. J. Monzon and L. L. Sanchez-Soto, Eur. J. Phys. {\bf 22}, 39 (2001),
 and the references contained in these papers.

\bibitem{bk02}
 S. Baskal and Y. S. Kim, Phys. Rev. {\bf E 66}, 06604 (2002), and
 the references contained in this paper.

\bibitem{azzam77}
R. A. M. Azzam and I. Bashara, {\it Ellipsometry and Polarized Light}
(North-Holland, Amsterdam, 1977);

\bibitem{wig39}
E. Wigner, Ann. Math. {\bf 40}, 149 (1939).

\bibitem{knp86}
Y. S. Kim and M. E. Noz, {\it Theory and Applications of the
Poincar\'e Group} (Reidel, Dordrecht, 1986).

\bibitem{bk03}
S. Baskal and Y. S. Kim, Phys. Rev. E (to be published), \\
or http://www.arXiv.org/abs/math-ph/0210056 (about lens optics and
contractions of Wigner's little groups).

\bibitem{monzon96}
J. J. Monz\'on and L. L. S\'anchez-Soto, Am. J. Phys. {\bf 64}, 156
(1996).

\bibitem{georg97}
I. J. Lalov and E. M.Georgieva, J. Mod. Opt., {\bf 44}, 265 (1997).

\bibitem{monzon00}
J. J. Monz\'on and L. L. S\'anchez-Soto, J. Opt. Soc. Am. A, {\bf 17},
1475 (2000); J. J. Monz\'on, T. Yonte, L. L. S\'anchez-Soto, and
J. Carinena, J. Opt. Soc. Am. A, {\bf 19}, 985 (2002).

\bibitem{hk88}
For the derivation of this formula based on Lorentz transformations,
see D. Han and Y. S. Kim, Phys. Rev. A {\bf 37}, 4494 (1988).

\end{thebibliography}
\end{document}